\documentstyle[twocolumn,aps,epsf]{revtex}
\makeatletter

                        \def\g{\gamma}       
                                            
         \def\s{\sigma}                   
           \def\k{{\bf k}}                 
\def\D{\Delta}           \def\W{\Omega}                                 
\def\S{{\cal S}}         \def\A{{\cal A}}          \def\WJ{W_{}}        
\def\N{{\cal N}}                  
\def\Den{{\cal D}}                                  
\def\DS{{\cal D}_{\rm S}}\def\RT{R_{\rm T}}        \def\RJ{R_{\rm J}}   
\def\VJ{V_{\rm J}}       \def\VT{V_{\rm T}}         
                 
\def\/{\over}            \def\<{\langle}           \def\>{\rangle}      
\def\us{\uparrow}        \def\ds{\downarrow}       \def\dmu{\delta\mu}  
\def\[{\left[}           \def\]{\right]}           
\def\({\left(}           \def\){\right)}

\def\Ispin{I_{\rm spin}} \def\Ipair{I_{\rm pair}}  \def\Iinj{I_{\rm inj}}
\def\Iqp{I_{\rm qp}}                                                    
\begin{document}
  \twocolumn[\hsize\textwidth\columnwidth\hsize\csname
  @twocolumnfalse\endcsname
\draft
\preprint{{\bf GH-13} (8$^{\rm th}$ Joint MMM-Intermag Conference, San Antonio)}
\widetext
%%%%%%%%%%%%%%%%%%%%%%%%%%%%%%%%%%%%%%%%%%%%%%%%%%%%%%%%%%%%%%%%%%%%%%%%%%%%%%%
%\title{ Spin current through Josephson junctions }
 \title{ \Large \bf Joule heating generated by spin current through Josephson junctions }
%%%%%%%%%%%%%%%%%%%%%%%%%%%%%%%%%%%%%%%%%%%%%%%%%%%%%%%%%%%%%%%%%%%%%%%%%%%%%%%
\author{S. Takahashi,\footnote{Electronic mail: takahasi@imr.tohoku.ac.jp}
T. Yamashita, T. Koyama, and S. Maekawa}
\address{Institute for Materials Research, Tohoku University, Sendai 980-8577, Japan}
\author{H. Imamura}
\address{Graduate School of Information Sciences, Tohoku University, Sendai 980-8579, Japan}
%-----------------------------------------------------------------------------
%%\date{\today}
\maketitle
\widetext
%%%%%%%%%%%
% Abstract 
%%%%%%%%%%%
{\begin{abstract}
%------------------------
We theoretically study the spin-polarized current flowing through
a Josephson junction (JJ) in a spin injection device.
When the spin-polarized current is injected from a ferromagnet (FM)
in a superconductor (SC),
the charge current is carried by the superconducting condensate
(Cooper pairs), while the spin-up and spin-down currents flow in the
equal magnitude but in the opposite direction in SC, because of no
quasiparticle charge current in SC.
This indicates that not only the Josephson current but also
the spin current flow across JJ at zero bias
voltage, thereby generating Joule heating by the spin current.
The result provides a new method for detecting the spin current
by measuring Joule heating at JJ.
%------------------------
\end{abstract} }
%% \pacs{PACS number: 73.40.Gk, 74.50.+r, 73.40.Rw}
\vskip 2.5pc]
\narrowtext

%%% Josephson effects, 74.50.+r
%%% Tunneling
%%% -interface structures, 73.40.G
%%% -superconductors, 74.50
%%% Tunnel junction devices, 85.30.M

%%\baselineskip 23pt
%%%%%%%%%%%%%%%%%%%%%%%%%%%%%%%
%\noindent{\bf I. INTRODUCTION}
%%%%%%%%%%%%%%%%%%%%%%%%%%%%%%%

Spin transport through a nonmagnetic metal has attracted much interest
in magnetic nanostructures.  In the tunnel junctions consisting of a
ferromagnet (FM) and a normal metal (N) or superconductor (SC), the
tunnel current driven from FM is spin-polarized \cite{meservey} and
creates a nonequilibrium spin population in N or SC
 \cite{johnson,johnsonS}.
Recently, there has been a number of experiment on suppression of
superconductivity by injection of spin-polarized electrons using tunnel
junctions of a high-$T_c$ SC and a ferromagnetic manganite
 \cite{vasko,dong,yeh}.

A double tunnel junction with a thin layer of SC sandwiched between
two FM electrodes is a unique system to investigate a novel
magnetoresistive effect caused by nonequilibrium spin population in SC.
When the thickness of SC layer is much smaller than the spin diffusion
length, the spin population depends strongly on the relative
orientation of magnetizations in FMs.
In the antiparallel alignment of magnetizations, the spin
population is accumulated in SC and reduces the superconducting gap
$\D$ with increase of tunneling currents, while in the parallel
alignment there is no such effect because of the absence of spin
accumulation in SC \cite{takahashi}.
In contrast, the spin current flows in SC in the parallel alignment,
but not in the antiparallel alignment \cite{takahashiPC}.
Although several methods for detecting the spin accumulation
has been presented, there is little for the spin current.
In this paper, we propose a new method for detecting the spin current
flowing in SC using a Josephson junction (JJ).   We show that the
spin current flows through JJ even at zero bias voltage ($\VJ=0$),
thereby generating Joule heating at JJ.   Since there
is no quasiparticle charge current across JJ at $\VJ=0$,
the Joule heating generated at $\VJ=0$ is a direct signature
of the spin current flowing in SC.

%%%%%%%%%%%%%%%
%\section{MODEL}
%%%%%%%%%%%%%%%

%%\topskip 4.8cm

We consider a spin injection device made of FM and SC separated by thin
insulating layers as shown in Fig.~1.  The central part of the junction
forms the Josephson junction, which is sandwiched between two FMs.
The left and right SC (FM) are made of the same SC (FM).
The magnetization of the left FM points up and that of the right FM is
either up or down.  
The applied bias current $\Iinj$ flows through the junctions of
resistances $\RJ$ and $\RT$ with the voltage drops $\VJ$ and $\VT$,
respectively.

We calculate the tunneling current using a phenomenological tunneling
Hamiltonian that describes the transfer of electrons from one electrode
to the other.  If SC is in the superconducting state, it is convenient
to rewrite the electron operators $a_{\k\s}$ in SC in terms of
quasiparticle operators $\g_{\k\s}$ appropriate to the superconducting
states, using the Bogoliubov transformation \cite{tinkham}
  %-------------------------------------------------------------
  \begin{eqnarray}  \nonumber                                   
   a_{\k\us}          =   u_\k\g_{\k\us}                        
                      + v_\k^*{\hat S}\g^\dagger_{-\k\ds},      
   \ \ \                                                
   a^\dagger_{-\k\ds} = - v_\k{\hat S}^\dagger\g_{\k\us}        
                      + u^*_\k\g^\dagger_{-\k\ds} ,             
  \end{eqnarray}                                                
  %-------------------------------------------------------------
where $|u_\k|^2 = {1\/2}\( 1+{\xi_\k/E_\k}\)$,
      $|v_\k|^2 = {1\/2}\( 1-{\xi_\k/E_\k}\)$,
${\hat S}$ is an operator which annihilates a Cooper pair, while
${\hat S}^\dagger$ creates one, and $E_\k=[{\xi_\k^2+\D^2}]^{1/2}$
is the quasiparticle dispersion of SC, $\xi_\k$ being the
one-electron energy relative to the chemical potential of the
condensate and $\D$ being the isotropic superconducting gap.

  \topskip 0cm

From the Fermi's golden rule result, the spin-dependent tunnel currents
$I_{i\s}$ across the $i$th junction are expressed as \cite{takahashi}
  %-----------------------------------------------------
\begin{mathletters}                                     
  \begin{eqnarray}                                      
    I_{1}^{\us}(\VT) = \({G_{1\us}/e}\) \[\N - \S_1 \], 
      \label{eq:I1-u} \\                                
    I_{1}^{\ds}(\VT) = \({G_{1\ds}/e}\) \[\N + \S_1 \], 
      \label{eq:I1-d} \\                                
    I_{2}^{\us}(\VT) = \({G_{2\us}/e}\) \[\N + \S_2 \], 
      \label{eq:I2-u} \\                                
    I_{2}^{\ds}(\VT) = \({G_{2\ds}/e}\) \[\N - \S_2 \]. 
      \label{eq:I2-d}                                   
  \end{eqnarray}                                        
\end{mathletters}                                       
  %-----------------------------------------------------
Here, $G_{i\s}$ ($i=1,2$) is the tunnel conductance
for electrons with spin $\s$ when SC is in the normal state.
The quantity $\N$ is given by the usual expression
  %---------------------------------------------
  \begin{equation}                              
    \N(\VT)  =  \int_{-\infty}^\infty  \DS(E)   
    \bigl[ f_0(E-{e\VT}) - f_0(E)\bigr] dE  ,   
     \label{eq:N}                               
  \end{equation}                                
  %---------------------------------------------
where $\DS(E)={\rm Re}\left[{|E|/\sqrt{E^2-\D^2}}\right]$ is the
normalized BCS density of states and $f_0(E)$ the Fermi function.
The quantity $\S_i$ is the quasiparticle spin density accumulated
in SC

%=======================================================
\begin{figure}                                          
  \epsfxsize=0.94\columnwidth                           
  \centerline{\hbox{\epsffile{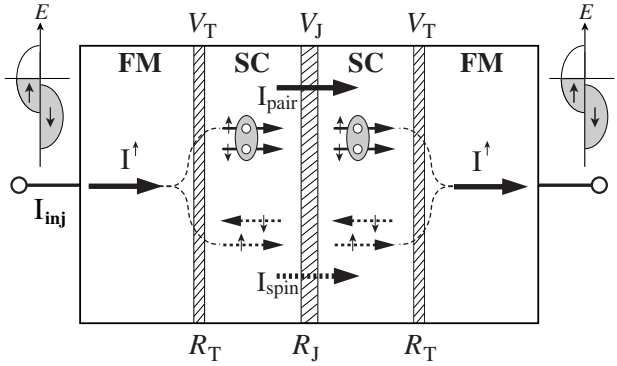}}}
 \vskip 0.5cm
  \caption{
%\small                                       
Spin injection device consisting of a Josephson junction
(SC-I-SC) which is sandwiched between two ferromagnets  
(FM).
 %% The central junction is a Josephson junction        
 %% with tunnel resistance $\RJ$.                       
The arrows indicate the injection current, pair current,
and quasiparticle current in SCs, and the Josephson current
(solid arrow) and spin current (dotted arrow) across JJ,
in the case of parallel alignment of magnetizations in FMs
whose spin polarization is 100 \%.
   }                                                    
\end{figure}                                            
%%=======================================================
\bigskip
\noindent
  %---------------------------------------------
  \begin{eqnarray}                              
    \S_i = {1\/2} \int_{-\infty}^\infty \DS(E)  
      \bigl[ f_{i\us}(E)-f_{i\ds}(E) \bigr] dE ,
     \label{eq:S}                               
  \end{eqnarray}                                
  %---------------------------------------------
where $f_{i\s}(E_\k)= \<\g^\dagger_{\k\s} \g_{\k\s}\>_i$ is the
distribution function of quasiparticles with energy $E_\k$ and
spin $\s$ in the $i$th SC.
In Eq.~(1), we neglected the charge imbalance ($Q^*$) since it has no
relevance to the spin-dependent effect \cite{takahashiPC}.

In the limit of vanishing spin-flip scattering in SC, the spin-up and
spin-down currents are treated as independent channels.  The conservation
of the total charge current $\Iinj=I_{i}^{\us} + I_{i}^{\ds}$ and the spin
current $\Ispin=I_{i}^{\us} - I_{i}^{\ds}$ across the left and right
junctions yields the relations
  %-------------------------------------
  \begin{eqnarray}                      
     \Iinj   &=& {1\/e\RT} [ \N- {1\/2}\(P_1\S_1-P_2\S_2\) ], \\
     \Ispin  &=& {1\/2e\RT}[ \(P_1+P_2\)\N - \(\S_1-\S_2\) ],
  \end{eqnarray}                        
  %-------------------------------------
where $1/\RT = G_{i\us} + G_{i\ds}$ and
$P_i = (G_{i\us}-G_{i\ds})/(G_{i\us}+G_{i\ds})$. $P_i$ represents
the degree of spin-polarization of FM.
In the parallel alignment, $P_1=P_2=P$ and $\S_1=-\S_2=\S$, so that
  %-------------------------------------
  \begin{eqnarray}                      
     \Iinj  = {1\/e\RT} \( \N- P\S \), \ \ \ \
     \Ispin = {1\/e\RT} \(P\N - \S \). 
  \end{eqnarray}                        
  %-------------------------------------
%%yielding the relation $\Iinj^{\rm(F)}=(G_T/e)\(1-P^2\)\N+P\Ispin^{\rm(F)}$.
In the antiparallel alignment, $P_1=-P_2=P$ and $\S_1=\S_2=P\N$, so that
  %-------------------------------------
  \begin{eqnarray}                      
     \Iinj   = (e\RT)^{-1} \( 1-P^2 \) \N, \ \ \ \
     \Ispin  = 0. 
  \end{eqnarray}                        
  %-------------------------------------

The tunnel current across JJ has the form
  %---------------------------------------------
  \begin{eqnarray}                              
   \Iinj=\Iqp(\VJ)+I_{\rm J1}(\VJ)\sin\varphi   
                  +I_{\rm J2}(\VJ)\cos\varphi,  
     \label{eq:I}                               
  \end{eqnarray}                                
  %---------------------------------------------
for a bias voltage $\VJ$ and a phase difference $\varphi$ of the gap
parameters on the two side of JJ.
  %% For a constant voltage $\VJ$, $\varphi=-2eVt/\hbar+\varphi_0$,
  %% where $e=|e|$ and $\varphi_0$ is a constant.
The first term describes the quasiparticle tunneling,
and second and third terms describe the phase coherent (Cooper pair)
tunneling.  The usual Josephson effect is associated with the
$\sin\varphi$ term.
Using the golden rule formula, we have the spin-dependent
tunnel current of quasiparticles
  %---------------------------------------------
  \begin{eqnarray}                              
    \Iqp^\s(\VJ) &=& {1\/2e \RJ}                
     \int_{-\infty}^\infty \DS(E)\DS(E+e\VJ)    
        \cr  &\times&                           
      \[ f_{1\s}(E) - f_{2\s}(E+e\VJ)  \] dE,   
     \label{eq:Iqpus}                           
  \end{eqnarray}                                
  %---------------------------------------------
with which the quasiparticle charge current $\Iqp$ and the spin current
$\Ispin$ across JJ are written as
  %---------------------------------------------
  \begin{eqnarray}                              
    \Iqp   &=& \Iqp^\us(\VJ) + \Iqp^\ds(\VJ),   
     \label{eq:Iqp}                             
     \\                                         
    \Ispin &=& \Iqp^\us(\VJ) - \Iqp^\ds(\VJ).   
     \label{eq:Ispin}                           
  \end{eqnarray}                                
  %---------------------------------------------
The phase coherent tunneling terms are obtained as
  %-----------------------------------------------------
  \begin{eqnarray}                                      
    I_{\rm J1} &=& {\D^2\/e \RJ}                        
     \int_{-\infty}^\infty dE                           
       { \DS(E)\DS(E+e\VJ) \/ |E(E+e\VJ)| }             
      \cr  &\times&                                     
     \sum_{j=1}^2 \[1 - f_{j\us}(|E|) - f_{j\ds}(|E|)\],
     \label{eq:IJ1}                                     
  \end{eqnarray}                                        
  %-----------------------------------------------------
  %-----------------------------------------------------
  \begin{eqnarray}                                      
    I_{\rm J2} &=& {\D^2\/e\RJ} \int_{-\infty}^\infty dE
       { \DS(E)\DS(E+e\VJ) \/ E(E+e\VJ) }               
        \cr &\times&                                    
         \sum_\s \[f_{2\s}(E+e\VJ) -f_{1\s}(E) \].      
     \label{eq:IJ2}                                     
  \end{eqnarray}                                        
  %-----------------------------------------------------
When the thickness of SC is much smaller than the spin diffusion
length, the distribution of quasiparticles is spatially uniform in SC.
Then, the distribution function $f_{i\s}$ is described by $f_0$, but
the chemical potentials of the spin-up and spin-down quasiparticles
are shifted oppositely by $\dmu_i$ from the equilibrium one to create
the spin density; 
  %-------------------------------------
  \begin{equation}                      
     f_{i\us}(E) = f_0(E - \dmu_i),     
        \ \ \ \                         
     f_{i\ds}(E) = f_0(E + \dmu_i).     
     \label{eq:fks}                     
  \end{equation}                        
  %-------------------------------------
The gap parameter $\D$ in SCs is determined by $f_{i\s}$ through
the BCS gap equation
  %---------------------------------------------
  \begin{eqnarray}                              
    \ln\[\D\/\D_0\]  + \int_{\D}^\infty %%\DS(E)
    {f_{i\us}(E) + f_{i\ds}(E)\/ \sqrt{E^2-\D^2}} dE =0,
    \label{eq:gap}                              
  \end{eqnarray}                                
  %---------------------------------------------
where $\D_0$ is the gap at $T=0$ in the equilibrium state ($\dmu_i=0$).

In the following we calculate the tunnel current at zero bias voltage
($\VJ=0$).   In this case, a DC Josephson current flows across JJ unless
the bias current exceeds the Josephson critical current $J_c=I_{\rm J}(0)$.
From Eq.~(\ref{eq:Ispin}), we notice that $\Ispin$ in the parallel
alignment becomes finite even at $\VJ=0$ if $\dmu_1$ and $\dmu_2$ take
nonzero values of different sign ($\dmu=\dmu_1=-\dmu_2$), while $\Ispin$
in the antiparallel alignment is zero because of $\dmu=\dmu_1=\dmu_2$.
The currents $I{\rm J2}$ and $\Iqp$ vanish at $\VJ=0$ irrespective of the
value of $\dmu_i$.  These facts indicate that the Josephson current as well
as the spin current flow across JJ at $\VJ=$ 0 in the parallel alignment,
while only the Josephson current flows in the antiparallel alignment.

%===============================================
\begin{figure}                                  
  \epsfxsize=0.86\columnwidth                   
  \centerline{\epsffile{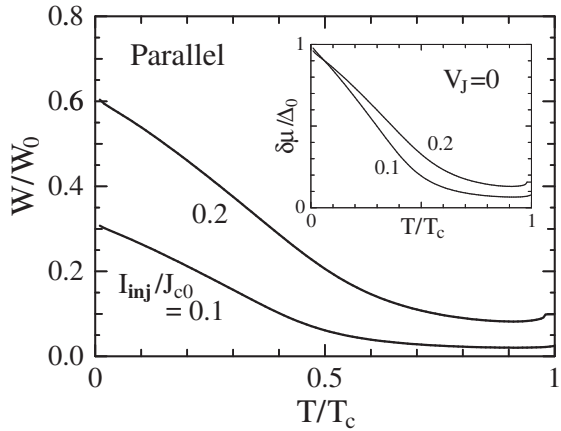}}       
  \vskip 0.5cm                                  
  \caption{
%\small                               
Joule heating $\WJ$ as a function of temperature
at zero bias ($\VJ=0$) in the parallel alignment
of magnetizations in FMs.  Inset shows half the 
spin splitting of the chemical potential of     
quasiparticles in SC.                           
   }                                            
\end{figure}                                    
%===============================================
\bigskip
\noindent

We consider, for simplicity, the case where FM is a half-metal with
100 \% spin polarization ($P=1$), in which the spin current is equal
to the injection current, $\Iinj=\Ispin$, because only the spin-up
electrons are injected into SC.   
Figure 1 shows how the current $\Iinj=I^\us$ is injected from the left
FM in SC and how the spin and charge currents flows in SCs and pass
through JJ in the parallel alignment.
In SC, the charge current is carried by the superconducting condensate
(Cooper pairs) alone; there is no charge current carried by quasiparticles
($\Iqp^\us + \Iqp^\ds=0$), so that the spin-up and spin-down
currents flow in the opposite direction with the value of
$\Iqp^\us = - \Iqp^\ds=I^\us/2$.   
In other words, the half of $I^\us$ is injected to be the spin-up current
$\Iqp^\us$ in SC, and the other half of $I^\us$ injected in SC is combined
with the backflow spin-down current $\Iqp^\ds$ to form the pair current
$\Ipair$.   It is remarkable that the spin current is able to pass
through JJ at zero bias ($\VJ=0$).
This is because the voltage drop across JJ is positive for the spin-up
current, while it is negative for the spin-down current, resulting in
no net voltage drop across JJ.
Therefore, the spin current as well as the Josephson current
flow through the JJ at $\VJ=0$, the magnitude of which is equal to
$\Iinj$ in the half-metallic case.

The most striking prediction of the present theory is that the spin
current across JJ is accompanied by Joule heating $\WJ$ at zero
bias voltage ($\VJ=0$).   The Joule heating $\WJ$ generated at $\VJ=0$
is given by $\WJ = \Ispin\(\dmu_1-\dmu_2\)/e$.
In the parallel alignment ($\dmu=\dmu_1=-\dmu_2$),
  %---------------------
  \begin{eqnarray}      
    \WJ = 2\Ispin\dmu/e,        
   \label{eq:W}         
   \end{eqnarray}       
  %---------------------
while in the antiparallel alignment $\WJ = 0$ because of $\Ispin=0$ and
$\dmu=\dmu_2=\dmu_1$, indicating that the Joule heating generated
by spin current has strong spin-dependence.
%%We self-consistently solve Eqs.~(\ref{eq:Ispin}) and (\ref{eq:gap})
%%with respect to $\D$ and $\dmu_S$, and calculate
%%the spin current and Joule heating.

Figure~2 shows the Joule heating $\WJ$ normalized to
$W_0=\D_0^2/e^2\RJ$ as functions of temperatures for the injection
current $\Iinj/J_{c0}=0.1$ and 0.2 in the parallel alignment \cite{dynes},
where $J_{c0}=\pi\D_0/(2e\RJ)$ is the Josephson critical current.
As the temperature is lowered, $\WJ$ increases monotonically.  This is
due to the increase of the spin-splitting of $\dmu$ (see the inset)
in the current bias mode.  A slight decrease of the critical
temperature and a depression of $\WJ$ near $T_c$ in the curve of
$\Iinj/J_{c0}=0.2$ is caused by the pair breaking effect by spin
accumulation in SC.  The power of Joule heating $\WJ$ generated
by spin current is estimated by evaluating the value of $W_0$,
since $\WJ$ is comparable to $W_0$.  If one uses the values of an
area resistance $\RJ\A=100$~$\W\mu$m$^2$ and $\D_0=0.3$~meV (Al),
we have $W_0/\A=90$~mW/cm$^2$ per area of JJ.  Therefore, $\WJ$ is
fairly large for observing the Joule heating experimentally.

In conclusion, we propose a new method for detecting the spin current
by measuring the Joule heating generated at JJ.  
This is attributed to the fact that the spin-up and spin-down currents
flow in the opposite direction in SC.
For measuring the Joule heating, it may be important for the tunnel
junctions to satisfy the condition $\RJ \gg \RT$, in order to make
the Joule heating $\WJ$ dominant compared with the injection power
($\Iinj^2\RT$).

\medskip

We are grateful to M.~Johnson and J.~S.~Moodera for valuable discussions.
This work is supported by a Grant-in-Aid for Scientific Research
Priority Area for Ministry of Education, Science and Culture of Japan,
NEDO Japan, and by the supercomputing facilities in IMR, Tohoku University.

%%%%%%%%%%%%%%%%%%%%%%%%%%%
%        References        
%%%%%%%%%%%%%%%%%%%%%%%%%%%
 
%%%%%%%%%%%%%%%%%%%%%%
%%%%%%%%%%%%%%%%%%%%%%

%%%%%%%%%%%%%%%%%%%%%%%%%%%
%     Figure Captions      
%%%%%%%%%%%%%%%%%%%%%%%%%%%


\begin{thebibliography}{10}
\bibitem{dynes}
To avoid a spurious logarithmic divergence arising from convolving
the BCS density of states with an isotropic gap in Eq.~(\ref{eq:Ispin}),
we introduce a small broadening in the BCS density of states using
Dynes's formula $\Den_S(E) = {\rm Re}[{(E-i\g)/\sqrt{(E-i\g)^2-\D^2 } }]$
with broadening parameter $\g$ much less than $\D$.
\end{thebibliography}
\end{document}